\documentclass[%
 reprint,
superscriptaddress,
amsmath,amssymb,
aps,
floatfix
]{revtex4-2}

\usepackage[utf8]{inputenc} 
\usepackage[T1]{fontenc}    
\usepackage{graphicx}       
\usepackage{dcolumn}        
\usepackage{bm}             
\usepackage{braket}         
\usepackage{booktabs}       
\usepackage{here}           
\usepackage[colorlinks=true, allcolors=blue]{hyperref} 
\usepackage{color}          
\usepackage{textcomp}       
\usepackage{comment}        
\usepackage{tabularx}       
\usepackage{mathptmx}
\usepackage{mathtools}
\usepackage{comment}

\begin{document}

%\preprint{APS/123-QED}

\title{Uncertainty quantification for proton-induced nucleon knockout reactions}

\author{T. Kondo}
\affiliation{Department of Physics, Kyushu University, Fukuoka 819-0395, Japan}

\author{C. Hebborn}
\affiliation{Universit\'e Paris-Saclay, CNRS/IN2P3, IJCLab, 91405 Orsay, France}
\affiliation{Facility for Rare Isotope Beams, Michigan State University, East Lansing, Michigan 48824, USA}
\affiliation{Department of Physics and Astronomy, Michigan State University, East Lansing, Michigan 48824, USA}

\author{Y. Chazono}
\affiliation{Department of Physics, Kyushu University, Fukuoka 819-0395, Japan}

\author{S. Tanaka}
\affiliation{Department of Physics, Kyushu University, Fukuoka 819-0395, Japan}

\author{F. M. Nunes}
\affiliation{Facility for Rare Isotope Beams, Michigan State University, East Lansing, Michigan 48824, USA}
\affiliation{Department of Physics and Astronomy, Michigan State University, East Lansing, Michigan 48824, USA}

\author{K. Ogata}
\email[]{ogata.kazuyuki.169@m.kyushu-u.ac.jp}
\affiliation{Department of Physics, Kyushu University, Fukuoka 819-0395, Japan}

\date{\today}
             
%%%%%%%%%%%%%%%%%%%%%%%%%%%%%%%%%%%%%%%%%%%%%%%%%%%%%%%%%%

\begin{abstract}
 \noindent
{\bf Background}:
Optical potentials are essential ingredients in the analysis of proton-induced nucleon knockout reactions. Their uncertainties can affect the calculated cross sections and the spectroscopic factors deduced from experimental data.
\\
{\bf Purpose}:
We investigate the propagation of uncertainties in nucleon optical potentials to $(p,2p)$ and $(p,pn)$ reaction observables and quantify their effects on the extracted spectroscopic factors.
\\
{\bf Methods}:
We perform distorted-wave impulse approximation (DWIA) calculations using 416 optical-potential parameters sampled from the KDUQ posterior distribution.
We consider six target nuclei from $^{12}$C to $^{208}$Pb at 197~MeV and compute $(p,2p)$ and $(p,pn)$ triple differential cross sections. We also calculate $^{40}$Ca$(p,2p)$ at 100~MeV. We estimate the relative uncertainties from the 68\% credible intervals of the triple differential cross sections at their peaks, choosing the higher-energy peak when two peaks are present.
\\
{\bf Results}:
We confirm that the distorted-wave impulse approximation provides a good description for the reaction mechanism, by comparing it to triple differential cross section and analysing power $A_y$ data for $^{40}$Ca$(p,2p)$ at 197~MeV. The relative uncertainty due to the optical potential is beam energy dependent: we find for $^{40}$Ca at 197~MeV, the relative uncertainty of the triple differential cross section is about 14\% for both $(p,2p)$ and $(p,pn)$, whereas it decreases to 10\% for $(p,2p)$ at 100~MeV. The optical potential of the incident proton gives the largest contribution to the uncertainty, when compared to the uncertainty coming from the optical potential in the exit channel. The results also show a dependence on the single-particle orbit of the knocked-out nucleon.
\\
{\bf Conclusions}:
The uncertainties in the spectroscopic factors deduced from $(p,2p)$ and $(p,pn)$ reactions at 197~MeV due to the optical potentials are typically about 15--20\% and 10\%, respectively, except for $^{208}$Pb$(p,2p)$ for which the uncertainty reaches about 30\% for $(p,2p)$ and 20\% for $(p,pn)$.

\end{abstract}
%%%%%%%%%%%%%%%%%%%%%%%%%%%%%%%%%%%%%%%%%%%%%%%%%%%%%%%%%%

\maketitle

%%%%%%%%%%%%%%%%%%%%%% SECTION.1 %%%%%%%%%%%%%%%%%%%%%%%%%
\section{introduction}\label{sec:intro}

Nuclear reactions are at the core of scientific programs at nuclear facilities worldwide \cite{crawford2024}. Reactions are versatile probes that enable the study of a wide range of properties of matter, from neutron skins to collectivity, from pairing to shell evolution at the limits of stability \cite{Brown_2025,ReviewAumannFAIR}. A crucial element for the analysis of all reaction measurements is a reliable model for the reaction dynamics. 

Historically, density distributions in nuclei were extracted through ($e,e'p$) reactions \cite{kramer2001}. However, until today, it remains impossible to produce electron targets. This limits the applicability of ($e,e'p$)  to studies with stable nuclei. Knockout ($p,pN$) reactions are the most promising alternative, since they are the closest probe to ($e,e'p$) and can be extended to rare isotopes. It has been demonstrated that ($p,2p$) and ($e,e'p$)  reactions provide consistent nuclear structure information \cite{Kaw18,GOMEZRAMOS2018511}.  Experimental studies that use proton-induced knockout abound: see for example, the studies exploring magicity in $^{52}$Ca \cite{Che19,Enc22} and $^{78}$Ni \cite{Ele19}.  Moreover, proton-induced knockout is standardly used in the production of rare isotopes (e.g. \cite{Sha17,Liu19}). Having a reliable reaction model that predicts proton-induced knockout cross sections with quantified uncertainties is a priority for our field.

Given the interest in proton-induced knockout reactions, there have been numerous studies dissecting the most commonly used reaction models used in interpreting these measurements, such as the study of the asymmetry in the parallel momentum distributions \cite{Oga15}, the dependence on binding energy \cite{GM18} and the role of pairing \cite{Yos18}. More importantly, the benchmarks of different reaction theories for ($p,pN$) have placed this probe on a very solid theoretical foundation \cite{Yos18,Gom20}. In those works, the Distorted Wave Impulse Approximation (DWIA), the method of choice in the present study, reproduced results from higher-complexity theories, such as the transfer-to-continuum method and the Faddeev framework \cite{Gom20}.
The DWIA method for ($p,pN$) requires nucleon-nucleus optical potentials as input, just as the other two methods benchmarked in \cite{Gom20}.

Optical potentials are widely used in nuclear reactions \cite{op2023}. Although much progress has been made toward extracting them directly from ab-initio theories, the community still relies on calibrating these potentials to nuclear data. While there have been a few studies involving  uncertainty quantification of the nucleus-nucleus optical potential (e.g. \cite{santos2026}), most of the work during the last decade has been focused on Bayesian uncertainty quantification for the nucleon-nucleus optical potential (e.g. \cite{lovell2021,Pru23,elm2026}). The KDUQ parameterization \cite{Pru23} is based on the global optical potential of Koning Delaroche \cite{KD03} and is valid for the high energies typically used in the ($p,pN$) studies. In previous work, the propagation of optical model uncertainties to reaction observables such as charge-exchange \cite{smith2024} and transfer \cite{PhysRevLett.131.212503,PhysRevLett.132.139901,HN25} using KDUQ has been studied. In this work, we extend the study to ($p,2p$) observables. This work presents a detailed analysis for $^{40}$Ca($p,2p$) at $E_p=100$ MeV and $197$ MeV but also extends the study to a wide range of stable closed shell nuclei  ($^{12}$C, $^{16}$O, $^{48}$Ca, $^{90}$Zr and $^{208}$Pb).

This paper is organized as follows. In Section II, we present the formalism, including a summary of the reaction theory (DWIA) and the global optical potential of choice (KDUQ). Numerical details are presented in Section III, followed by the results and discussion in Section IV.  Finally, Section V contains the summary and conclusions.

%%%%%%%%%%%%%%%%%%%%%% SECTION.2 %%%%%%%%%%%%%%%%%%%%%%%%%
\section{Formalism}
\label{sec:form}

\subsection{Distorted-wave impulse approximation}
\label{sec:DWIA}

We adopt the DWIA formalism described in
Refs.~\cite{Wak17,Oga24}. Here we briefly introduce
its basic structure. The incident proton is labeled
as particle 0, the struck nucleon as particle $N$,
and the two outgoing nucleons as particles 1 and 2.
The target and residual nuclei are denoted by
A and B, respectively.

To illustrate the formalism, we consider a simplified
case in which the spin-orbit parts of the distorting
potentials are neglected. The reduced transition
matrix is then given by
\begin{align}
\mathfrak{T}_m ={}&
\int d{\bm R}\,
\chi_1^{(-)*}({\bm R})
\chi_2^{(-)*}({\bm R})
\nonumber\\
&\times\chi_0^{(+)}({\bm R})
e^{-i(m_N/m_{\rm A}){\bm K}_0\cdot{\bm R}}
\psi_{nljm}({\bm R}),
\label{eq:dwia_tmat}
\end{align}
where $\chi_i^{(+)}$ and $\chi_i^{(-)}$ are the distorted
waves of particle $i$ satisfying the outgoing and incoming
boundary conditions, respectively, and $\psi_{nljm}$ is the single-particle (s.p.) wave
function of the struck nucleon. Here, $n$, $l$,
and $j$ are the radial quantum number, orbital
angular momentum, and total angular momentum,
respectively, and $m$ is the projection of $l$.
${\bm K}_0$ is the wave number vector of the
incident proton in the center-of-mass frame of
the reaction system, and $m_N$ and $m_A$ are
the masses of particle $N$ and nucleus A,
respectively. The exponential factor accounts
for the recoil effect.

With a factorization approximation for the
$NN$ transition amplitude, the triple differential cross section (TDX) is expressed as
\begin{equation}
\frac{d^3\sigma}
{dT_1^{\rm L}d\Omega_1^{\rm L}d\Omega_2^{\rm L}}
=S_{nlj}\mathfrak{K}^{\rm L}
\frac{d\sigma_{NN}}{d\Omega_{NN}}
\frac{1}{2l+1}
\sum_m|\mathfrak{T}_m|^2,
\label{eq:dwia_tdx}
\end{equation}
where $\mathfrak{K}^{\rm L}$ is the kinematical
factor, $d\sigma_{NN}/d\Omega_{NN}$ is the
differential cross section for free $NN$ scattering,
and $S_{nlj}$ is the spectroscopic factor
for the s.p. orbit of interest.
The explicit form of $\mathfrak{K}^{\rm L}$ is
given in Ref.~\cite{Oga24}.

In the actual calculations, we include the
spin-orbit parts of the distorting potentials
and use the spin-dependent NN transition amplitudes.
The full DWIA formulation, including the calculation
of the analyzing power $A_y$, is given in
Ref.~\cite{Oga24}.

\subsection{Uncertainty quantification with Koning-Delaroche optical potential}
\label{sec:KDUQ}

  The distorted waves  $\chi_i$ entering in the T-matrix are computed from  nucleon-nucleus optical potentials.  In this work, we use the democratic version of  the KDUQ global optical potential \cite{Pru23}, which has the same functional form as the  Koning-Delaroche global parametrization \cite{KD03}. Its 46 parameters encode energy, mass and isospin dependence of the potentials and were calibrated by a Bayesian inference against a large corpus of elastic, analyzing power and integrated cross-section data. Based on this corpus of  data, the authors of Refs.\,\cite{KD03, Pru23} restrict the validity to target masses  $A \geq 24$ and beam energies $1$ keV $<E<200$ MeV. 
  
  In this work, we use the 416   parameter sets of the democratic posterior distribution, provided in the Supplemental Material of Ref. \cite{Pru23}. We  propagate them through the DWIA model to the quasifree scattering observables,  while preserving correlations between the optical potentials: the potentials for particles 0, 1, and 2 are generated from the same KDUQ parameter set. At each kinematic point, the 68\% (95\%) credible intervals (CIs) are computed from the resulting 416 predictions  as the smallest intervals containing 68\% (95\%) of the quasifree scattering cross sections.

  To quantify and compare the magnitude of the uncertainty of  quasifree scattering observables, we define the relative half-width $\varepsilon_{68\%} $ of the 68\% CI $[\sigma_{in}^{68\%}(T_{1,max}^{\rm L}),\sigma_{max}^{68\%}(T_{1,max}^{\rm L})]$ at the peak of the cross section (corresponding to   the energy $T_{1,max}^{\rm L}$ of particle~1 in the laboratory frame) as
\begin{equation}
    \varepsilon_{68\%} = \frac{\sigma_{max}^{68\%}(T_{1,max}^{\rm L}) -\sigma_{avg}^{68\%}(T_{1,max}^{\rm L})  }{\sigma_{avg}^{68\%}(T_{1,max}^{\rm L}) }
\end{equation}
with 
$\sigma_{avg}^{68\%}(T_{1,max}^{\rm L}) = [\sigma_{max}^{68\%}(T_{1,max}^{\rm L}) +\sigma_{min}^{68\%}(T_{1,max}^{\rm L})]/2$
A similar definition was used for transfer reactions in Ref. \cite{HN25}.
  
%%%%%%%%%%%%%%%%%%%%%% SECTION.3 %%%%%%%%%%%%%%%%%%%%%%%%%
% Numerical inputs
\section{Numerical inputs}
\label{sec:num}

\begin{table}%[ht]
\centering
\caption{Reaction systems and kinematical conditions adopted
in the calculations. All energies are given in MeV and
angles in degrees.}
\label{tab:systems}
\begin{tabular}{ccccccc}
\toprule
System
& $T_0^{\rm L}$
& $S_N$
& Orbit
& $\theta_1^{\rm L}$
& $\theta_2^{\rm L}$
& $T_{1,max}^{\rm L}$  \\
%& (MeV) & (MeV) & & (deg.) & (deg.) & (MeV) \\
\midrule
$^{12}\mathrm{C}(p,2p)$
& 197 & 16.0  & $0p_{3/2}$ & 39.0 & 39.4 & 116 \\
$^{12}\mathrm{C}(p,pn)$
& 197 & 18.72 & $0p_{3/2}$ & 39.0 & 39.4 & 117 \\
\addlinespace
$^{16}\mathrm{O}(p,2p)$
& 197 & 12.1  & $0p_{1/2}$ & 39.4 & 39.4 & 113 \\
$^{16}\mathrm{O}(p,pn)$
& 197 & 15.66 & $0p_{1/2}$ & 39.4 & 39.4 & 114 \\
\addlinespace
$^{40}\mathrm{Ca}(p,2p)$
& 100 & 8.328 & $0d_{3/2}$ & 42.2 & 42.2 & 62 \\
$^{40}\mathrm{Ca}(p,2p)$
& 197 & 8.328 & $0d_{3/2}$ & 42.2 & 42.2 & 123 \\
$^{40}\mathrm{Ca}(p,pn)$
& 197 & 15.64 & $0d_{3/2}$ & 42.2 & 42.2 & 123 \\
\addlinespace
$^{48}\mathrm{Ca}(p,2p)$
& 197 & 15.8  & $1s_{1/2}$ & 39.4 & 39.4 & 90 \\
$^{48}\mathrm{Ca}(p,pn)$
& 197 & 9.95  & $0f_{7/2}$ & 39.4 & 39.4 & 133 \\
\addlinespace
$^{90}\mathrm{Zr}(p,2p)$
& 197 & 8.4   & $1p_{1/2}$ & 41.4 & 43.0 & 113 \\
$^{90}\mathrm{Zr}(p,pn)$
& 197 & 11.97 & $0g_{9/2}$ & 41.4 & 43.0 & 139 \\
\addlinespace
$^{208}\mathrm{Pb}(p,2p)$
& 197 & 8.0   & $2s_{1/2}$ & 42.0 & 42.0 & 95 \\
$^{208}\mathrm{Pb}(p,pn)$
& 197 & 7.37  & $2p_{1/2}$ & 42.0 & 42.0 & 110 \\
\bottomrule
\end{tabular}
\end{table}

We consider the reactions listed in Table~\ref{tab:systems}. We follow the Madison convention and assume coplanar kinematics for the three outgoing particles. The azimuthal angles $\phi_i$ for the two nucleons are set to $0^\circ$ or $180^\circ$. The nucleon emitted at $\phi_i=0^\circ$ ($180^\circ$) is called particle 1 (2), whereas the incident proton is called particle 0. The kinetic energy and the scattering angle of particle $i$ in the laboratory frame are denoted by $T_i^{\rm L}$ and $\theta_i^{\rm L}$, respectively. In Table~\ref{tab:systems}, the incident energy $T_0^{\rm L}$, the separation energy $S_N$ of the struck nucleon, its single-particle (s.p.) orbit, $\theta_1^{\rm L}$, and $\theta_2^{\rm L}$ are shown. In what follows, we discuss the UQ results for observables at $T_{1,max}^{\rm L}$.%; we adopt the CIs evaluated at {\lq\lq}$T_1^{\rm L}$ for UQ'' as representative results (see below).

\begin{table}%[ht]
\centering
\caption{Parameters of the bound-state potentials.
The radius and diffuseness parameters are given in fm and
the spin-orbit strengths in MeV.}
\label{tab:bound_potential}
\begin{tabular}{lcccccc}
\toprule
Nucleus
& $p$ or $n$
& $r_{\rm c}$
& $a_{\rm c}$
& $r_{\rm s}$
& $a_{\rm s}$
& $V_{\rm s}$ \\
\midrule
$^{12}\mathrm{C}$  & $p,n$ & 1.35  & 0.65  & 1.35  & 0.65  & 8.0 \\
$^{16}\mathrm{O}$  & $p,n$ & 1.37  & 0.65  & 1.37  & 0.65  & 7.1 \\
$^{40}\mathrm{Ca}$ & $p,n$ & 1.30  & 0.65  & 1.30  & 0.65  & 7.1 \\
$^{48}\mathrm{Ca}$ & $p$   & 1.228 & 0.65  & 1.228 & 0.65  & 8.3 \\
$^{90}\mathrm{Zr}$ & $p$   & 1.32  & 0.65  & 1.32  & 0.65  & 7.5 \\
\bottomrule
\end{tabular}
\end{table}

The s.p. potential for the nucleon in the target nucleus is given by
\begin{equation}
V(R)=-V_{\rm c} f_{\rm c}(R)+V_{\rm s}\frac{\lambda_\pi^2}{R}\frac{d}{dR}f_{\rm s}(R)\,{\bm l}\cdot {\bm \sigma}+V_{\rm cl}(R),
\label{vsp}
\end{equation}
\begin{equation}
f_{\rm x}(R)=
\left[ 1+ \exp\left([R-r_{\rm x} B^{1/3}]/a_{\rm x}\right) \right]^{-1},
\quad {\rm x}=\text{c or s} 
\label{vso2}
\end{equation}
where $B$ is the mass number of the residual nucleus and $\lambda_\pi=\sqrt{2}$~fm is the pion Compton wavelength. The Coulomb interaction $V_{\rm cl}(R)$ is evaluated by assuming a uniformly charged sphere with a radius of $r_{\rm c} B^{1/3}$ for the residual nucleus; this potential is zero for a neutron. The parameters used are listed in Table~\ref{tab:bound_potential}. For the neutron in $^{48}$Ca and $^{90}$Zr, and both proton and neutron in $^{208}$Pb, we employ the s.p. potential of Bohr and Mottelson~\cite{BM69}. The depth $V_{\rm c}$ of the central part is determined to reproduce the $S_N$ values shown in Table~\ref{tab:systems}.

To account for nonlocality, we multiply the distorted waves and the  s.p. wave function by the Perey factor~\cite{PB62} with a nonlocality range of 0.85~fm. We include partial waves up to the orbital angular momentum $L=40$ for the distorted waves of particles 0, 1, and 2. For evaluating the transition matrix, we carry out the integration over $R$ up to 15~fm.
For all the numerical calculations in this study, we adopt the computer code {\sc pikoe}~\cite{Oga24} with a slight modification to read the optical potential parameters.

%%%%%%%%%%%%%%%%%%%% SECTION.4 %%%%%%%%%%%%%%%%%%%%%%
\section{Results and Discussion}
\label{sec:res}

\begin{figure}%[!htpb]
\includegraphics[width=0.95\hsize]{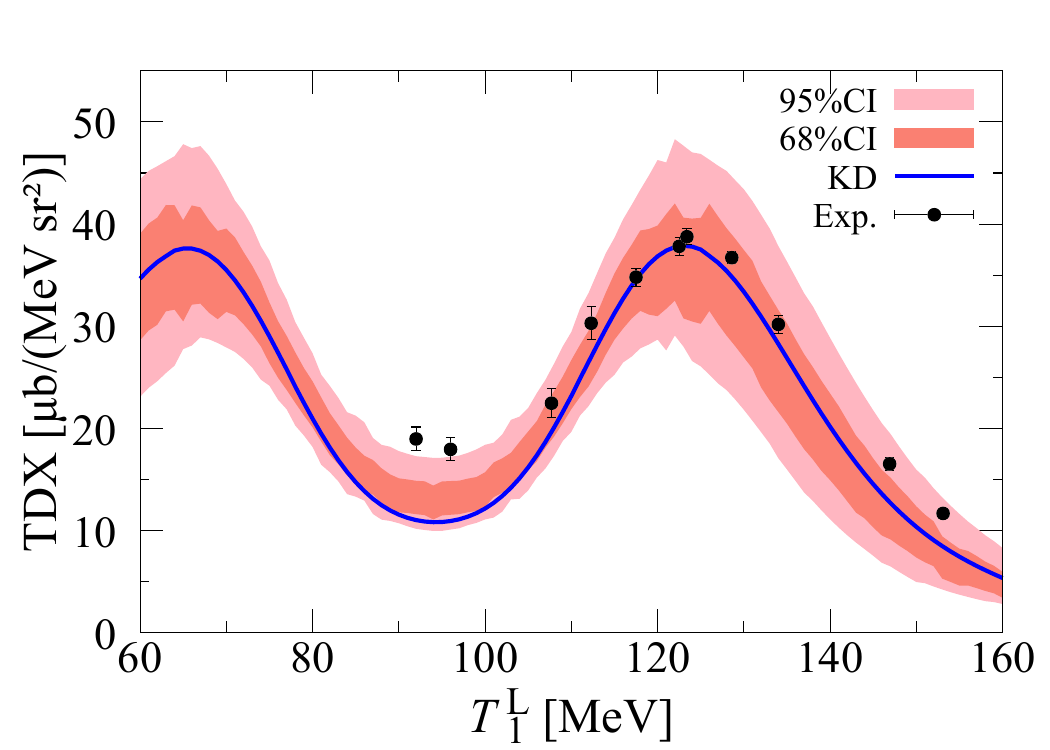}
\caption{
TDX for $^{40}$Ca($p,2p$)$^{39}$K (g.s.) at 197~MeV
as a function of $T_1^{\rm L}$.
The blue line represents the result with the original
KD potential parameters, while the darker and lighter
shaded bands represent the 68\% and 95\% CIs,
respectively. The black points are the experimental data~\cite{Nor23}.
}
\label{fig:40Cap2p197}
\end{figure}

Figure~\ref{fig:40Cap2p197} displays the TDX for $^{40}$Ca($p,2p$) at 197~MeV as a function of the energy $T_1^{\rm L}$ of particle~1 in the laboratory frame. In all the calculations discussed in the present study, we consider the reaction processes populating the ground states of the residual nuclei. The blue line is the result calculated with the original KD potential parameters. The darker (inner) and lighter (outer) shaded bands represent the 68\% and 95\% CIs associated with the uncertainties in the optical potential parameters. In all the calculations shown in Fig.~\ref{fig:40Cap2p197}, a spectroscopic factor of 2.492 is used~\cite{Yos22}. %To draw the bands, we adopted the prescription given in Ref.~\cite{HN25}. 
One sees that the blue line is within the 95\% CI band but for some $T_1^{\rm L}$ lies at the edge of the 68\% band.  This is consistent with the comparison of the reaction cross sections computed with KD and KDUQ (see Fig. 9 of Ref. \cite{Pru23}). We have estimated the relative uncertainty $\epsilon_{68\%}$ of the calculated TDX at the right-hand peak to be 0.14. This is close to the 10-15\% estimation of the optical potential uncertainties mentioned in Ref.~\cite{Wak17}. It should be noted that in the dip region, a correction for the finite momentum and solid-angle widths of the measurement is needed to compare the calculation with the data~\cite{Wak17}. To save computation time, we do not apply this correction in the present study.

\begin{figure}%[!htpb]
\includegraphics[width=0.95\hsize]{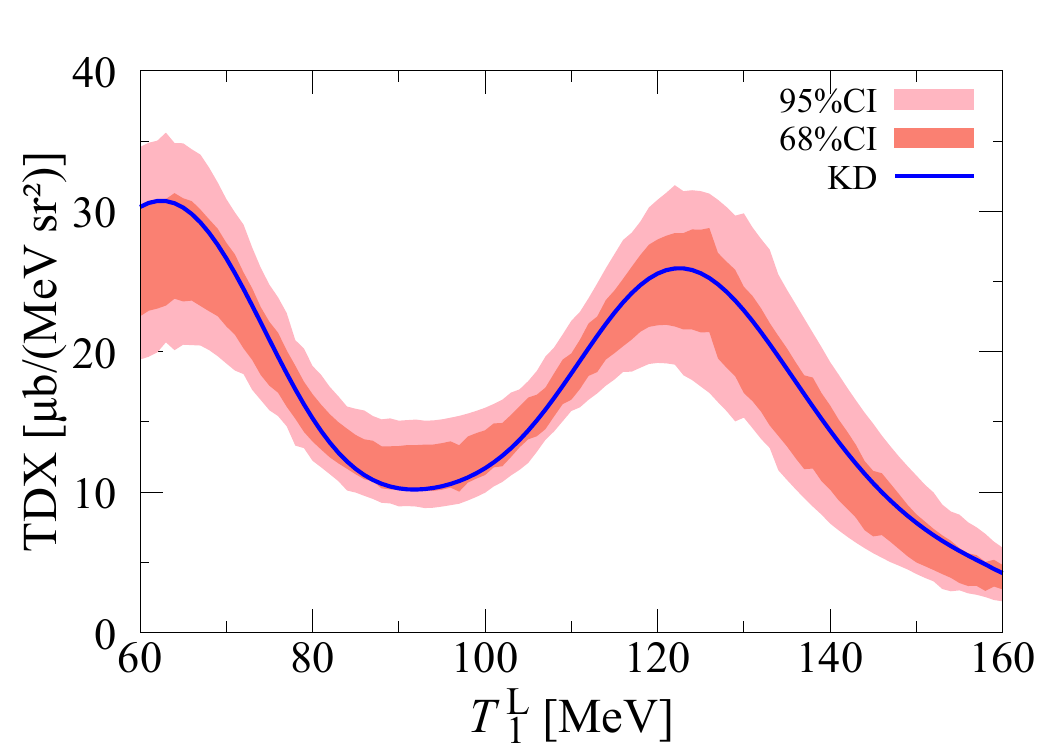}
\caption{Same as Fig.~\ref{fig:40Cap2p197}, but for
$^{40}$Ca($p,pn$)$^{39}$Ca (g.s.).}
\label{fig:40Cappn197}
\end{figure}

Next we show in Fig.~\ref{fig:40Cappn197} the TDX for $^{40}$Ca($p,pn$) at 197~MeV. The features of the result are almost the same as those in Fig.~\ref{fig:40Cap2p197}; $\epsilon_{68\%}$ is found to be 0.14. We have also analyzed $^{40}$Ca($p,2p$) at 100~MeV and obtained  $\epsilon_{68\%}=0.10$, suggesting that lower beam energies are associated with smaller uncertainties.

\begin{figure}%[!htpb]
\includegraphics[width=0.95\hsize]{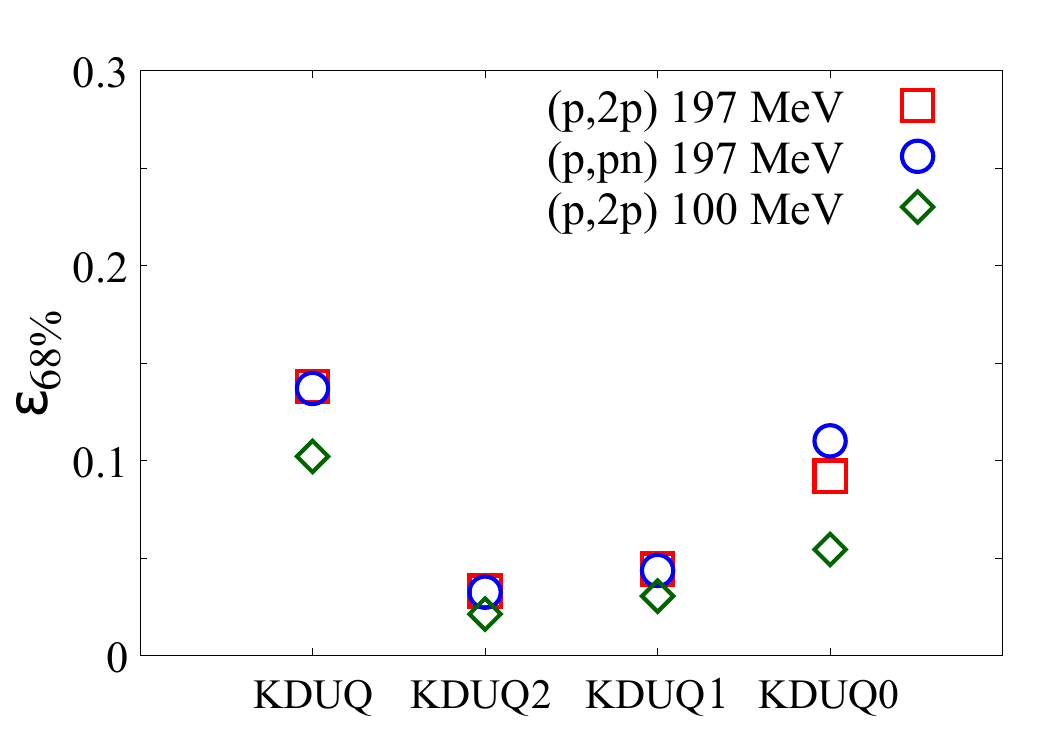}
\caption{
Comparison of $\epsilon_{68\%}$ for
$^{40}$Ca($p,2p$) at 100 and 197~MeV and
$^{40}$Ca($p,pn$) at 197~MeV.
KDUQ denotes UQ for all three optical potentials,
while KDUQ$i$ ($i=0,1,2$) denotes UQ for
particle $i$ only.
}\label{fig:summary40Ca}
\end{figure}

The leftmost set of results (KDUQ) in Fig.~\ref{fig:summary40Ca} summarizes $\epsilon_{68\%}$ discussed above. Besides KDUQ, three sets of results (KDUQ2, KDUQ1, and KDUQ0) are shown. Here, KDUQ$i$ ($i=0,1$ or 2) means that the UQ is performed only for the optical potential of particle $i$. KDUQ0 dominates over KDUQ1 and KDUQ2, which can be understood as follows. The energy of particle 0 is very close to the maximum energy of the KD parametrization, so its potential may be less well constrained. The slight difference in $\epsilon_{68\%}$ between KDUQ2 and KDUQ1 may also be understood from the energies of particles 1 and 2. For the 197~MeV cases, we evaluate $\epsilon_{68\%}$ at $T_1^{\rm L}\sim 120$~MeV, which corresponds to $T_2^{\rm L}\sim 70$~MeV. The smaller value of $\epsilon_{68\%}$ for KDUQ2 may suggest that the KD potential is better constrained at lower energies. A similar energy dependence of the uncertainty on the beam energy has been observed in the preceding study on nucleon transfer reactions~\cite{HN25} and in total, reaction, and elastic cross sections \cite{xhgl-d14x}.

\begin{figure}%[!htpb]
\includegraphics[width=0.95\hsize]{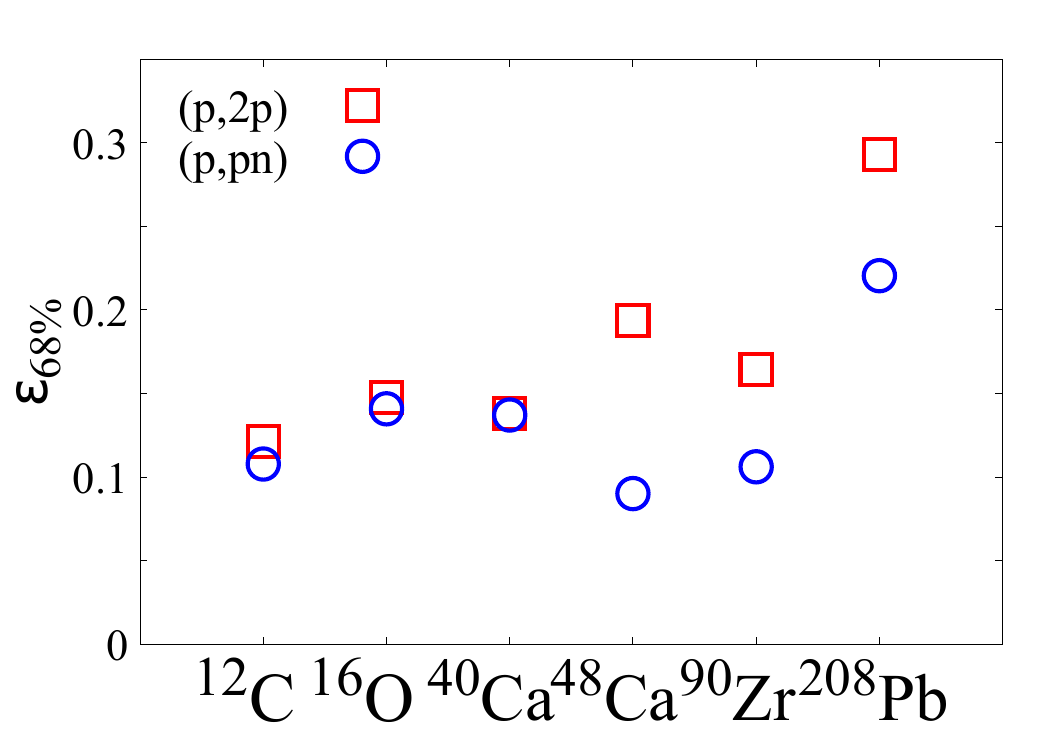}
\caption{
Relative uncertainties $\epsilon_{68\%}$ of the TDXs
for $(p,2p)$ and $(p,pn)$ reactions on six target nuclei
at 197~MeV. The values are evaluated at the energies
$T_1^{\rm L}$ listed in Table~\ref{tab:systems}.
}
\label{fig:summary6nuclei}
\end{figure}

Next we show in Fig.~\ref{fig:summary6nuclei} $\epsilon_{68\%}$ for $(p,2p)$ and $(p,pn)$ TDXs at 197~MeV, for several closed-shell nuclei across a wide range of mass. For a non-$s$-wave nucleon knockout process, the TDX has two major peaks. In that case, we have evaluated $\epsilon_{68\%}$ at  $T_{1,max}^{\rm L}$. For $^{12}$C, $^{16}$O, and $^{40}$Ca, $\epsilon_{68\%}$ is 10--15\% for both $(p,2p)$ and $(p,pn)$. It should be noted that $^{12}$C and $^{16}$O are outside the range of applicability of the KD potential. The results in Fig.~\ref{fig:summary6nuclei} suggest that the KD potentials for these nuclei are rather well constrained, as is the KD potential for $^{40}$Ca. In fact, it was shown in Ref.~\cite{Nor23} that DWIA calculations with the KD potential  reproduce well the TDXs for $^{12}$C$(p,2p)$ and $^{16}$O$(p,2p)$ measured at 197~MeV with a level of agreement comparable to that for heavier nuclei up to $^{90}$Zr.

$^{48}$Ca shows a clear difference between $(p,2p)$ and $(p,pn)$. This can be attributed to the difference in the s.p. orbits. The knocked-out proton and neutron are in the $1{\rm s}_{1/2}$ and $0{\rm f}_{7/2}$ orbits, respectively. The s.p. wave function with a larger $l$ tends to extend farther outward, which makes the knockout reaction more peripheral. Because the optical potential near the origin is difficult to constrain, $\epsilon_{68\%}$ tends to be larger (smaller) when the knockout reaction probes the inner (outer) region of nuclei. The results for $^{90}$Zr and $^{208}$Pb can be understood in the same manner.

Moreover, $^{208}$Pb is an exception, in that $\epsilon_{68\%}$ exceeds 0.2 and almost reaches 0.3 for $(p,2p)$. As shown in Fig.~31 in Ref.~\cite{Wak17}, the absorption effect is extremely large and about 97\% of the flux is lost to unobserved channels. This suggests that the cross sections are very sensitive to the distorted waves, and thus to the KD potential.
One aspect not discussed so far is the isospin dependence of the KDUQ uncertainties. For nuclei with $N \neq Z$, previous studies \cite{smith2024} suggest that these uncertainties might grow.

To summarize the results in Fig.~\ref{fig:summary6nuclei}, the relative uncertainty in the TDXs at 197~MeV, and hence in the spectroscopic factors to be deduced from $(p,2p)$ and $(p,pn)$ reactions, is about 15--20\% for $(p,2p)$ and about 10\% for $(p,pn)$, except for $^{208}$Pb.

\begin{figure}%[!htpb]
\includegraphics[width=0.95\hsize]{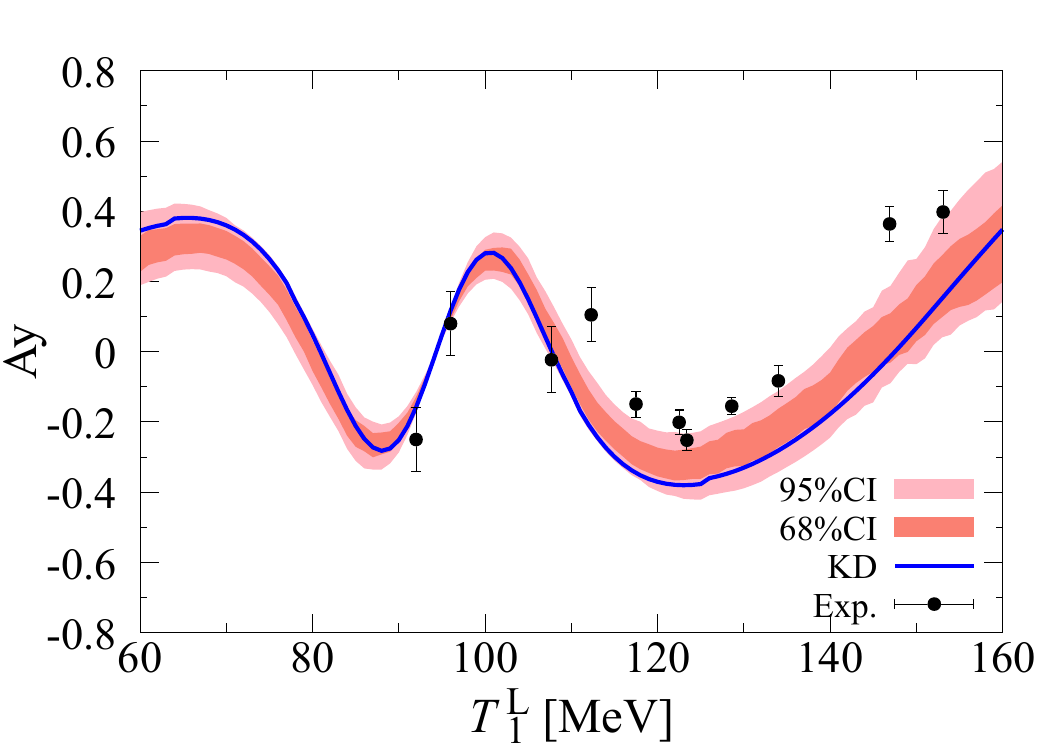}
\caption{Same as Fig.~\ref{fig:40Cap2p197},
but for the analyzing power $A_y$.}
\label{fig:40Cap2pAy}
\end{figure}

Finally, we show in Fig.~\ref{fig:40Cap2pAy} the UQ result for the analyzing power $A_y$ for $^{40}$Ca$(p,2p)$ at 197~MeV. In Ref.~\cite{Nor23}, it was argued that the DWIA calculation with the EDAD1 set of the Dirac phenomenology reproduces $A_y$ better than that with the KD potential.
As seen from the figure, the 95\% CI band covers most of
the experimental data, although some deviations remain
at higher $T_1^{\rm L}$.

%%%%%%%%%%%%%%%%%%%%%% SECTION.5 %%%%%%%%%%%%%%%%%%%%%%%%%
\section{summary}
\label{sec:sum}
We have investigated the uncertainties in $(p,2p)$ and $(p,pn)$ reaction observables originating from the nucleon optical potentials within the DWIA framework. We employed the KDUQ global optical model and propagated its uncertainties to  quasifree  reaction observables. The calculations were performed at 197~MeV for six target nuclei from $^{12}$C to $^{208}$Pb, and also at 100~MeV for $^{40}$Ca$(p,2p)$.

For $^{40}$Ca at 197~MeV, the relative uncertainty $\epsilon_{68\%}$ of the TDX was found to be 0.14 for both $(p,2p)$ and $(p,pn)$. The uncertainty decreases to 0.10 for $(p,2p)$ at 100~MeV. We found that the optical potential of the incident proton contributes more to the uncertainty than the optical potential of the outgoing wave. The results for the other nuclei suggest that the uncertainty depends on the s.p. orbit of the knocked-out nucleon and possibly on the isospin asymmetry of the target nuclei \cite{smith2024,HN25,Wak17}. In particular, $\epsilon_{68\%}$ becomes as large as 0.3 for $^{208}$Pb$(p,2p)$. We have also shown that the 95\% CI band for the analyzing power $A_y$ of $^{40}$Ca$(p,2p)$ covers most of the experimental data.

Our study focused on the parametric uncertainties associated with the optical potentials used to generate the distorted wave in the calculation of quasifree scattering observables.
Our results suggest that the uncertainty in the spectroscopic factors deduced from $(p,2p)$ and $(p,pn)$ reactions at 197~MeV  is typically about 15--20\% and 10\%, respectively, except for $^{208}$Pb. These values should be regarded as  a lower bound on the total uncertainty budget, since the uncertainties associated with the nucleon-nucleon operator in the T-matrix and the model inaccuracies were not quantified in this study.  Quantifying the former will be the subject of future work.  Finally, this study  considered the KDUQ global optical model, while other uncertainty-quantified models  are starting to emerge. It will be interesting  to compare how the uncertainties in quasifree scattering observables obtained with KDUQ compared to the ones obtained with these other parametrizations \cite{elm2026}, which  rely on different phenomenological forms, different statistical models and/or a different corpus of data in the calibration.

%%%%%%%%%%%%%%%%%%%%%%%%%%%%%%%%%%%%%%%%%%%%%%%%%%%%%%%%%%%
\begin{acknowledgments}
This work was supported in part by JST ERATO Grant No. JPMJER2304, Japan, and by Grants-in-Aid of the Japan Society for the Promotion of Science (Grant No. JP25K07302). The computation was carried out with the computer facilities at the Research Center for Nuclear Physics, the University of Osaka. This project also received financial support from the CNRS through the AIQI-IN2P3 project. The work of F.M.N. was in part supported by the U.S. Department of Energy grant DE-SC0021422 and National Science Foundation CSSI program under award No. OAC-2004601 (BAND Collaboration).

\end{acknowledgments}


%\nocite{*}
%\bibliographystyle{apsrev4-2}% Produces the bibliography via BibTeX.
\bibliography{ko2026}

\end{document}